# $NO_2$ and Humidity Sensing Characteristics of Few-layer Graphene

Anupama Ghosh†‡, Dattatray J. Late†, L. S. Panchakarla†, A. Govindaraj†‡ and

C. N. R. Rao*†‡

† Chemistry and Physics of Materials Unit, Jawaharlal Nehru Centre for advanced Scientific Research, Jakkur, P.O., Bangalore 560064, India

‡ Solid Sate and structural Chemistry Unit, Indian institute of science, Bangalore 560012, India

**Abstract:**

Sensing characteristics of few-layer graphenes for $NO_2$ and humidity have been investigated with graphene samples prepared by the thermal exfoliation of graphitic oxide (EG), conversion of nanodiamond (DG) and arc-discharge of graphite in hydrogen (HG). The sensitivity for $NO_2$ is found to be highest with DG. Nitrogen-doped HG (n-type) shows increased sensitivity for $NO_2$ compared to pure HG. The highest sensitivity for humidity is observed with HG. The sensing characteristics of graphene have been examined for different aliphatic alcohols and the sensitivity is found to vary with the chain length and branching.

# 1. Introduction:

Gas sensor characteristics of various nanostructures have been investigated in the last few years [1-3]. Thus, many metal oxide nanostructures show good sensing characteristics for gases such as $NO_2$, $NH_3$, hydrocarbons and ethanol [1-10]. Carbon nanotubes (CNTs) are known to exhibit fast response and high sensitivity for detection of small concentrations of toxic gases at room temperature. Semiconducting CNTs can be used for detecting very small concentration of $NH_3$, $NO_2$ and other gases [11-15].

Recent studies on the interaction of graphene with gas molecules have indicated that it can act as a good sensor [16-24]. Schedin and co-workers [22] have shown that the increase in the charge carrier concentration induced by gas molecules adsorbed on the surface of graphene can be used to fabricate sensitive gas sensors. Based on theoretical investigations on the adsorption of gas molecules on single-layer graphene as well as on graphene nanoribbons, it has been predicted that the doping in carbon nanostructures may improve the sensitivity [25-28]. Ao and co-workers have applied density functional theory to show that aluminium doped graphene can be used as a good detector for carbon monoxide [29]. Functionalized graphite nanostructures are able to sense traces of pollutant gases such as $NO_2$ [30]. Water vapor sensing characteristics of reduced graphene oxides has been studied [31]. Reduced graphene oxide is also shown to be good sensor for toxic vapors [32]. Spin-coated hydrazine functionalized graphene dispersions are able to detect $NO_2$, $NH_3$, and 2,4-dinitrotoluene [33].

We considered it important to examine gas sensing characteristics of few-layer graphene prepared by different methods. For this purpose, we have prepared graphene by the thermal exfoliation of graphitic oxide (EG), conversion of nanodiamond (DG) and by arcing graphite rods in a hydrogen atmosphere (HG) [34-37]. We have also prepared nitrogen- and boron-doped graphene samples (N-HG and B-HG) to study the effect of doping on the gas sensing characteristics. We have studied sensing of $NO_2$ and humidity by all the above graphene samples. We have also examined the sensing characteristics of EG for different aliphatic alcohols.

## 2. Experimental technique:

Graphene was prepared by four different methods, namely the exfoliation of graphitic oxide (EG), conversion of nanodiamond (DG), and arc discharge of graphite in a hydrogen atmosphere (HG). To prepare EG, graphitic oxide was prepared by reacting graphite powder with a mixture of concentrated nitric acid and sulfuric acid with potassium chlorate at room temperature for five days and the thermal exfoliation of graphitic oxide was carried out in a long quartz tube at $1050^oC$ under an Argon atmosphere [34]. Thermal conversion of nanodiamond to graphene was carried out at $1650^oC$ in a helium atmosphere to obtain DG [35]. To prepare HG, direct current arc-discharge of graphite evaporation was carried out in a water cooled stainless steel chamber filled with a mixture of hydrogen and helium with the proportion $H_2$ (200 torr)-He (500 torr) with the discharge current in the 100-150 A range and maximum open circuit voltage of 60 V [37]. Boron doped graphene (B-doped HG) was prepared by carrying out arc-discharge using a boron-stuffed graphite electrode (3 at% boron) in the presence of $H_2$ (200 torr) and He (500 Torr). Nitrogen doped graphene (N-doped HG) was prepared by carrying out arc-discharge of graphite electrodes in the presence of $H_2$ (200 torr), He (200 torr) and $NH_3$ (300 torr).

The as-synthesized graphene samples were characterized by X-ray Diffraction (Cu K$\alpha$ radiation), transmission electron microscopy (TEM) (JEOL JEM 3010), field emission scanning electron microscopy (FESEM) (Nova Nanosem 600), atomic force microscopy (AFM) (CP 2 atomic force microscope), and Raman spectroscopy (Labraman-HR) using an He–Ne laser (632.81 nm). Brunauer-Emmett-Teller surface areas were measured using a Quantachrome Autosorb-1 instrument. The surface areas were measured at 77 K using $N_2$ as the adsorbate.

The sensing devices were prepared as follows, 300 nm thick gold film was deposited on a glass substrate by thermal evaporation to make source and drain with a 15 μm separation between the electrodes. Graphene samples were dispersed in methanol using ultrasonication. 5μL of the dispersion was dropped on to the electrodes by dielectrophoresis.

Gas sensing properties were measured using a home-built computer-controlled characterization system consisting of a test chamber, sensor holder, a Keithley multimeter-2700, a Keithley electrometer-6517A, mass flow controllers and a data acquisition system. The test gas was mixed with $N_2$ to achieve the desired concentration and the flow rate was maintained using mass flow controllers. By monitoring the output voltage across the sensor, the resistance of the sensor in dry air or in the test gas can be measured. The resistance of the graphene samples increased on contact with $NO_2$, while the resistance decreased in contact with water and alcohol vapours. The sensitivity (response magnitude) S was determined as the ratio $\Delta R/R_{air}$, where $\Delta R$ is the difference of resistance of the graphene sample in the presence of the test gas and in dry air, $R_{air}$ is the resistance of the samples in dry air. The resistance of the sensors prepared by us, based on graphene was in the range of 0.1–15 kΩ. The response time is defined as the time required for the resistance to reach 90% of the equilibrium value after the test gas is injected and recovery time is taken as the time necessary for the sensor to attain a resistance 10% above the original value in air. The controlled humidity environments were achieved using saturated aqueous solutions of LiCl, $MgCl_2$, $K_2CO_3$, NaBr, KI, NaCl, KCl and $K_2SO_4$ in a closed glass vessel at an ambient temperature of 25°C which yielded approximately 11.3, 32.8, 43.1, 57.5, 68.8, 75.3, 84.3 and 97.3 % relative humidity (RH) respectively. These RH levels were independently monitored by using a hygrometer (Keithley 6517A). Then a Keithley multimeter was used to measure the change of the sensor resistance in the testing circuit. To measure alcohol sensing characteristics of EG sample, $N_2$ gas was bubbled at a constant flow rate (100 sccm) through a gas bubbler containing different alcohols to get alcohol vapours with alcohol concentration around 200 ppm in the test chamber.

## 3. Results and discussion:

We have characterized the graphene samples by a variety of techniques. Figures 1(a) and (b) show typical transmission electron micrographs and Raman spectra of thermally exfoliated graphene (EG) and arc-discharge graphene (HG) samples. The TEM image of EG shows existence for 4-5 layers of graphene while bi- and tri-layer graphenes are mostly present in HG. The Raman spectra show the presence of the D, G and 2D bands in all the samples [36]. The intensity of 2D band is greater in HG than in EG. Analysis of the (002) reflections in the X-ray diffraction patterns of the grapheme samples shows that EG and DG samples possess between 3 and 6 graphene layers, while HG, N-doped HG and B-doped HG samples possess between 2 and 3 layers only and that was further verified by AFM cross-section height profile analysis. The BET surface areas of EG and DG were high (1260 $m^2$/g and 930 $m^2$/g respectively) but that of HG and doped HG samples were rather low (~ 400 $m^2$/g). Typical low-magnification FESEM images of dielectrophoretically deposited graphene samples (EG and DG), taken at magnifications of 1000x and 1500x respectively between two gold electrodes are shown in Figures 2(a) and (b) respectively. In the inset of Figure 2(a), we show a higher magnification image (20000x) of EG, revealing the sheet-like morphology of graphene.

We have measured the current–voltage (I-V) characteristics for all the graphene samples at 1000 ppm $NO_2$ and different relative humidity. Figure 3(a) shows the typical Ohmic behaviour of N-doped HG in air and in $NO_2$. Figure 3(b) shows I-V characteristics of DG at 4% and 84% relative humidity. These I-V characteristics demonstrate that the graphene samples can be used for sensing these vapours.

Figures 4(a) and (b) show typical gas sensing characteristics of the graphene samples for different concentrations of $NO_2$. We find the highest sensitivity with DG, the value reaching 65%. The response time is quite reasonable with both HG and DG, the values being around 15 min, but it is high with EG (~50 min). Rapid response and recovery times are found when the molecular adsorption occurs on low-energy binding sites. The sensing characteristics of graphene for $NO_2$ are fully reversible on heating the samples to 150 $^o$C to remove the adsorbed gases.

We have examined the effect of doping graphene on the $NO_2$ sensing characteristics of HG prepared by arc-discharge in hydrogen. The sensitivity of HG decreases on boron doping and increases significantly on nitrogen doping. Figure 4(c) shows the sensing characteristics of the nitrogen-doped HG sample for different concentrations of $NO_2$. It appears that n-type graphene is a better sensor for $NO_2$ as it is an electron withdrawing molecule. The response times with the boron- and nitrogen-doped HG samples are 15 and 50 minutes respectively. In figure 5(a), we show the variation of sensitivity for the graphene samples exhibiting the highest sensitivity. We see that the value of the sensitivity increases with $NO_2$ concentration and is satisfactory beyond 100 ppm. We have carried out sensitivity measurements for $NO_2$ over repeated cycles and obtained reproducible results. For practical applications, however, it may be necessary to anneal the sensor-device and remove the absorbed $NO_2$ after each cycle.

The sensitivity of the three graphene samples for humidity was measured at 25$^o$ C and 60$^o$C. Typical sensing characteristics are shown in figures 6(a) and (b). High sensitivity for humidity is found with HG, the value reaching 80%. The response time varies between 3 and 5 min for the three graphene samples. In figure 5(b), we show the typical variation of sensitivity with the relative humidity in the case of HG at $25^o$ C and $60^o$C. The sensitivity increases with % RH as expected. The sensitivity is satisfactory above 20% RH.

We have also examined the sensing characteristics of EG for different aliphatic alcohols. In the case of normal aliphatic alcohols, the sensitivity varies with the chain length in the order ethanol > n-propanol > n-butanol as shown in figure 7(a). The sensitivity depends on branching in the case of isomeric butyl alcohols with t-butanol > iso-butanol > n-butanol as shown in Figure 7(b).

## 4. Conclusions:

In conclusion, thick film sensors prepared with few-layer graphene show satisfactory sensing characteristics for $NO_2$ and $H_2O$. Graphene prepared by nanodiamond conversion shows the best sensitivity for $NO_2$ although it does not have the highest surface area. It may be because the surface of nanodiamond-converted graphene (DG) is least functionalized compared to EG and HG. Nitrogen-doped graphene shows enhanced sensitivity for $NO_2$ since the latter is an electron-withdrawing molecule. HG prepared by arc-discharge in a $H_2$ atmosphere shows the best sensitivity for humidity. Since water is an electron donor molecule, it appears that HG which has no oxygen functional groups shows the best sensing characteristics. It is interesting that the sensing characteristics of EG for aliphatic alcohols depends on the chain length and branching.

**Figure captions**

**Figure 1.** TEM images and Raman spectra of (a) EG and (b) HG samples.

**Figure 2.** FESEM images of dielectrophoretically deposited (a) EG and (b) DG between two gold electrodes.

**Figure 3.** I-V characteristics of (a) N-doped HG (at 25 °C) in air and in 1000 ppm of $NO_2$ and (b) DG (at 25 °C) in 4% and 84% RH.

**Figure 4.** Gas sensing characteristics of (a) DG (b) HG, and (c) N-doped HG for 1000, 500, 100 and 50 ppm of $NO_2$.

**Figure 5.** (a) Variation of sensitivity of DG and N- doped HG for $NO_2$ with the concentration of $NO_2$. (b) Variation of sensitivity of HG with relative humidity at 25°C and 60°C.

**Figure 6.** The change in sensitivity of (a) EG and (b) HG accompanying the dynamic switch between dry air (4% RH) and 84 % RH.

**Figure 7.** Alcohol sensing characteristics of EG (all alcohols at 200 ppm) at 25 °C: (a) methanol, ethanol, n-propanol and n-butanol (b) n-butanol, iso-butanol and t-butanol.

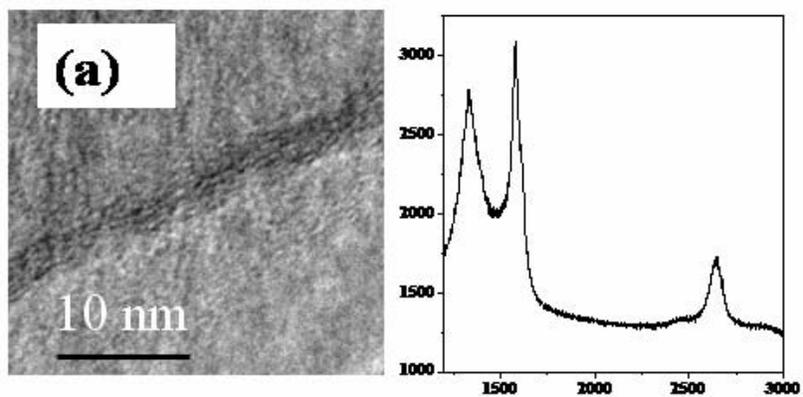
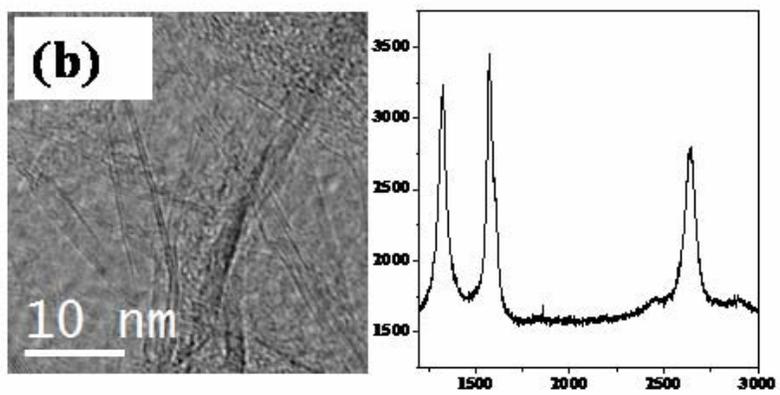

**Figure 1**

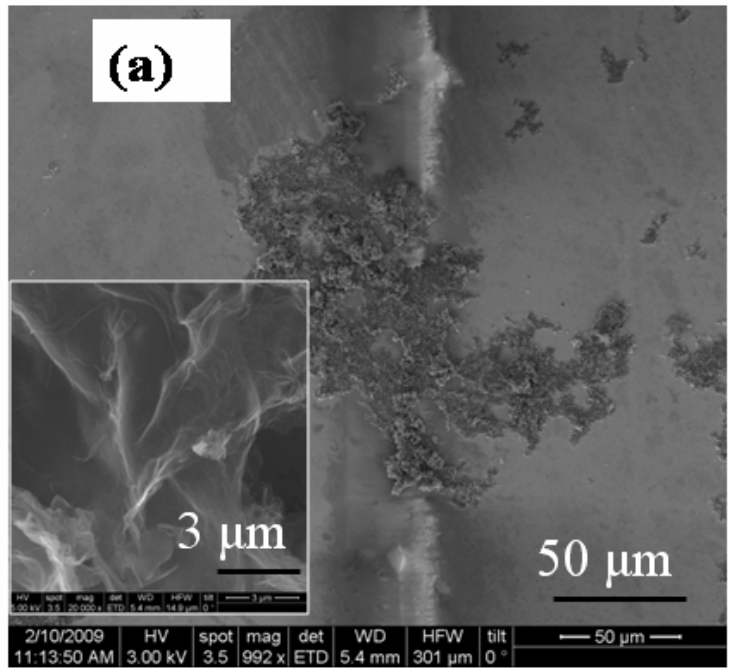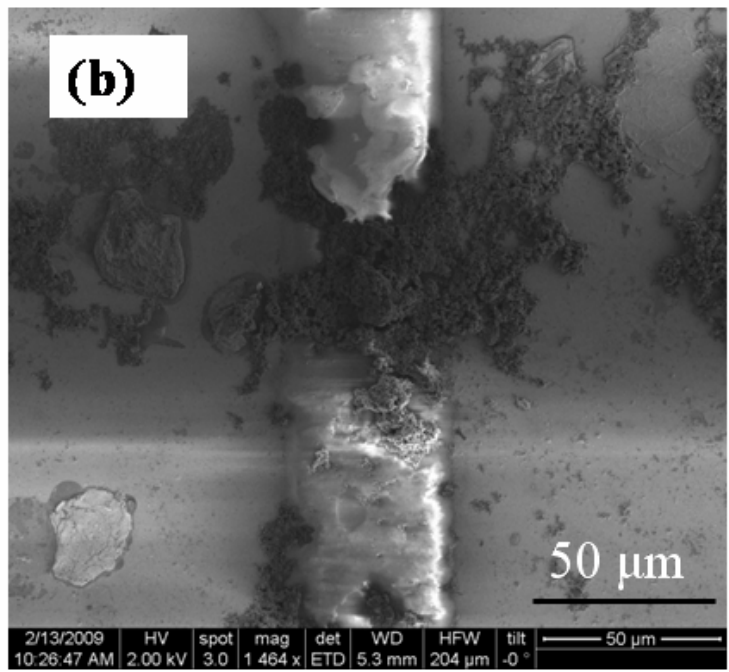

**Figure 2**

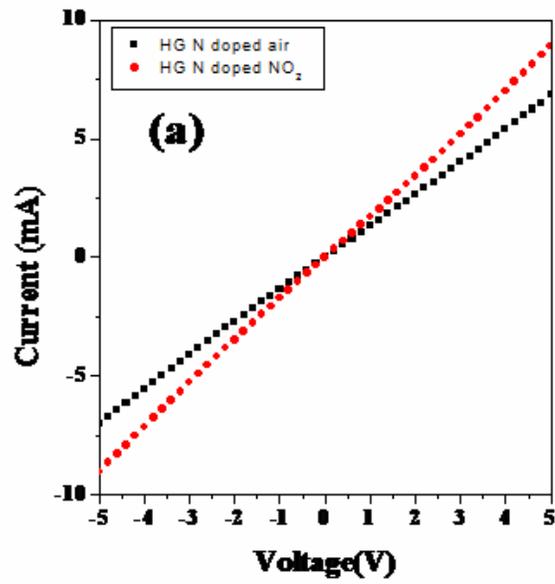

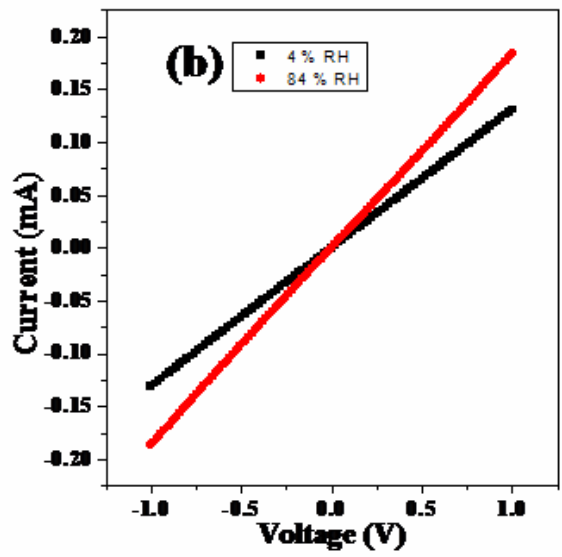

**Figure 3**

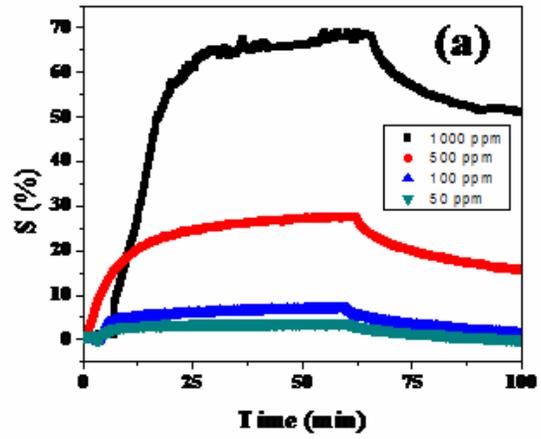
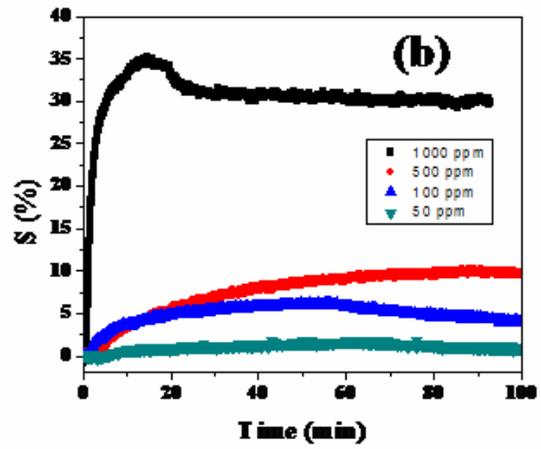
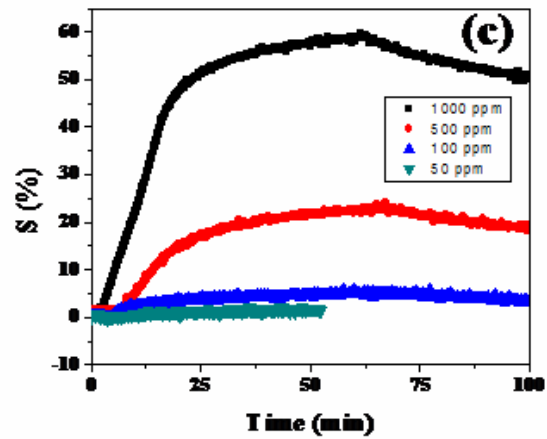

**Figure 4**

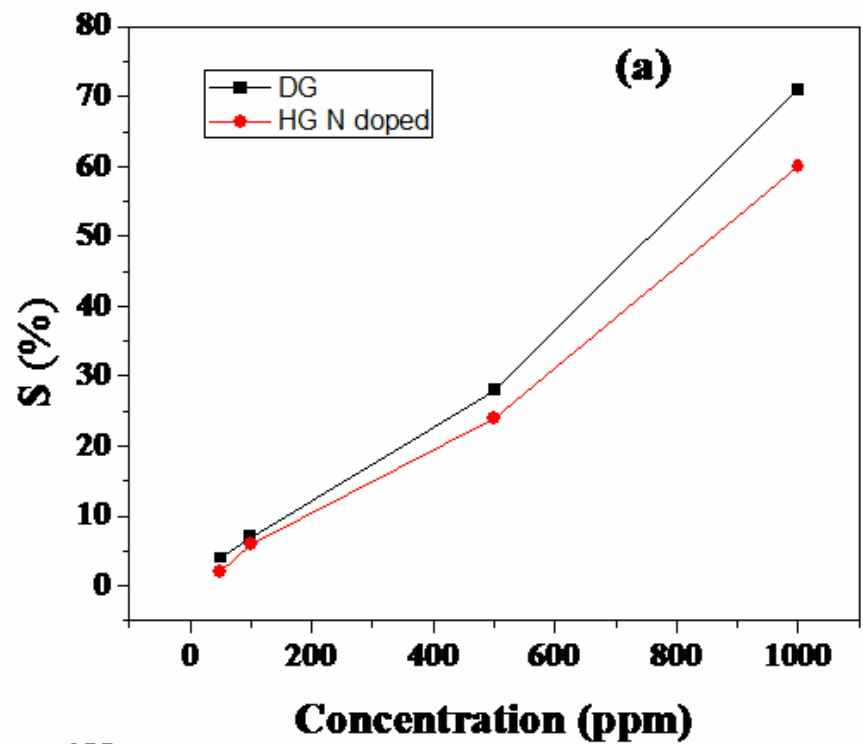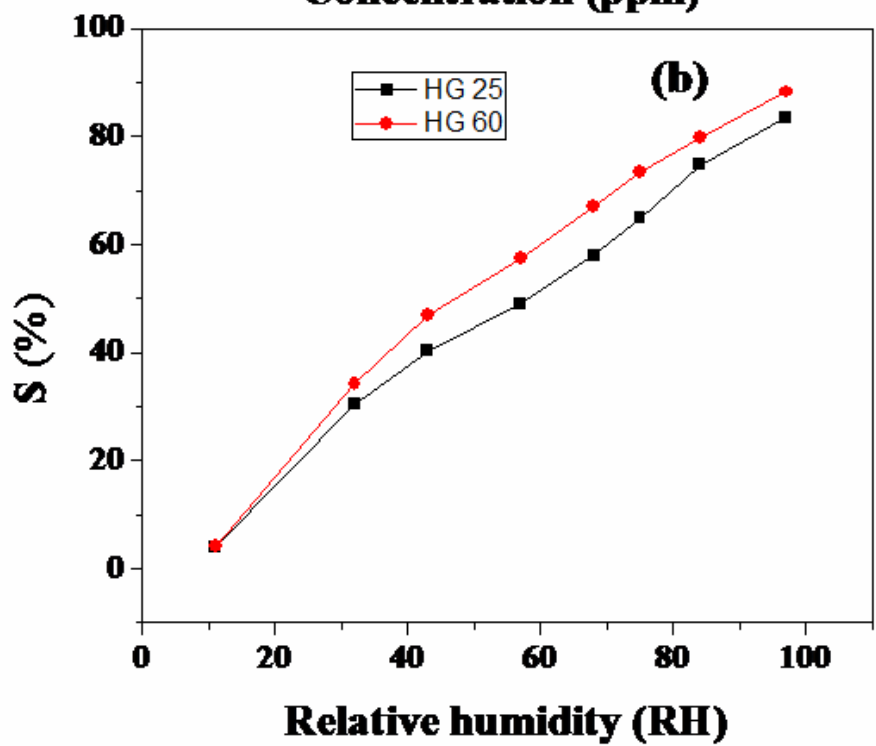

Figure 5

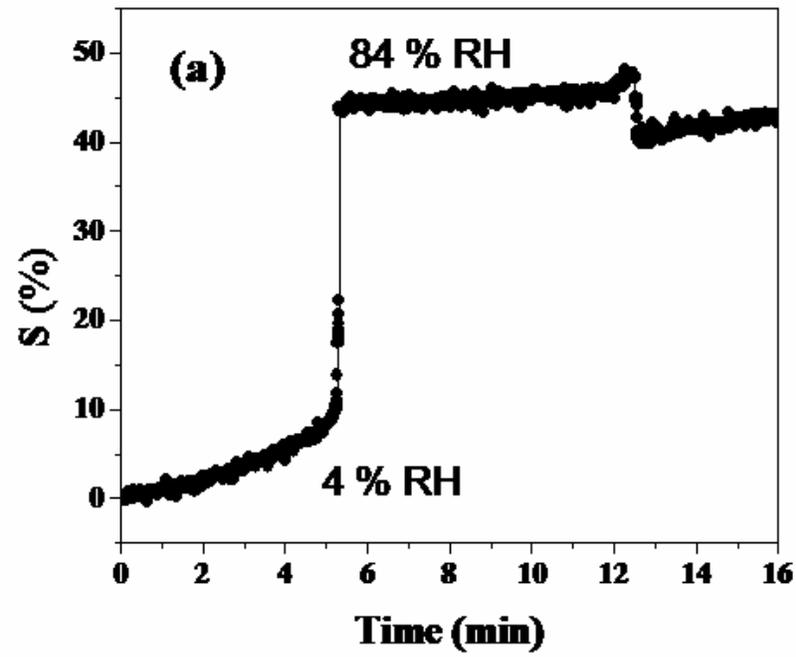

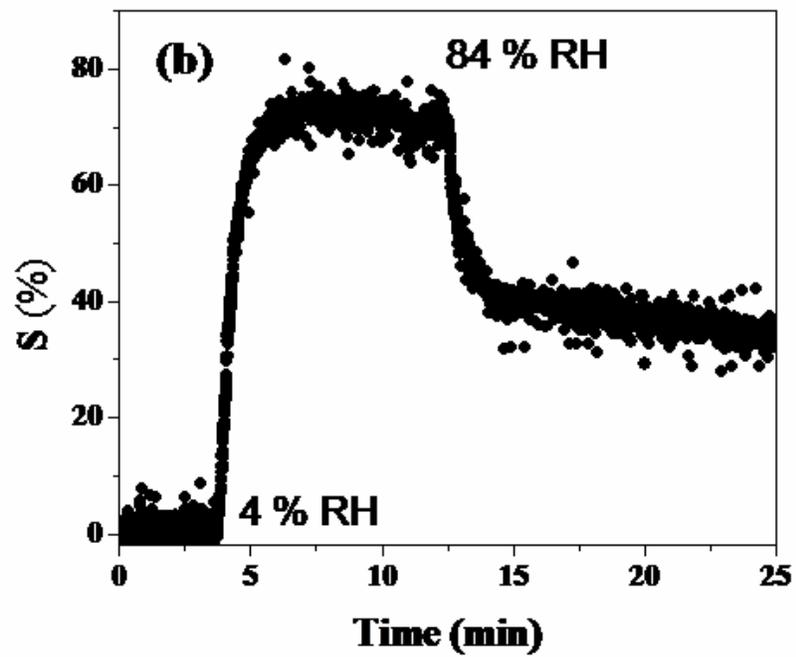

**Figure 6**

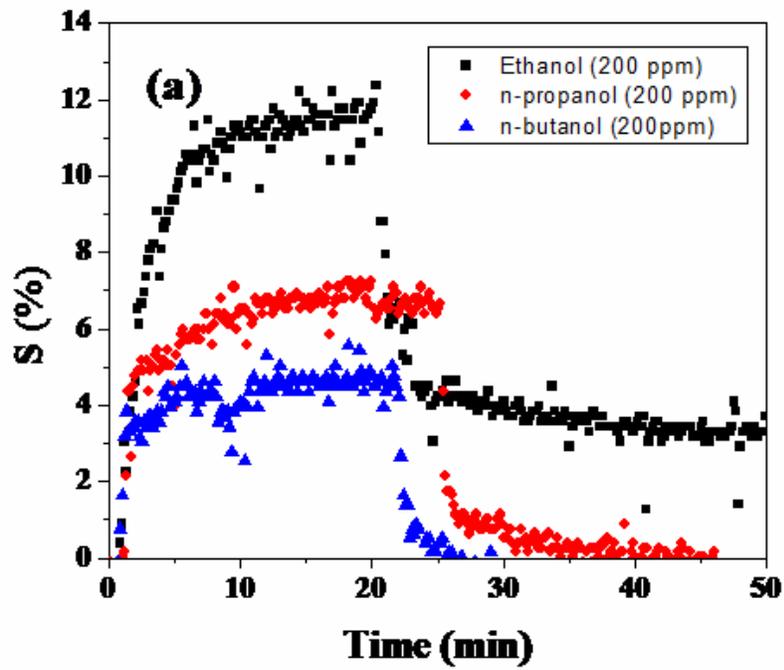

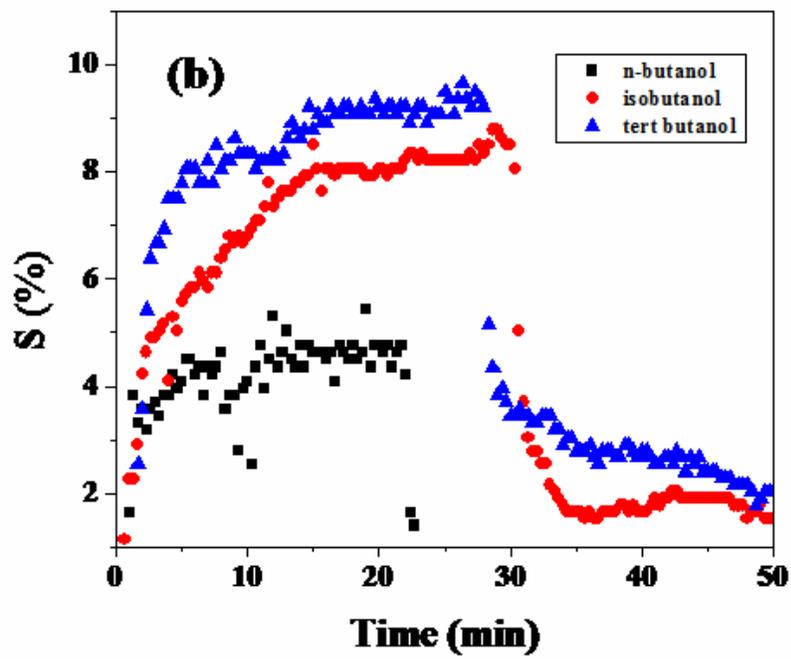

**Figure 7**